\documentclass[12pt]{article}
\usepackage{amsfonts,amsmath}
\usepackage{amssymb}
\usepackage{latexsym}

\def\beq{\begin{equation}}
\def\eeq{\end{equation}}
\def\bea{\begin{eqnarray}}
\def\eea{\end{eqnarray}}
\def\ba{\begin{array}}
\def\ea{\end{array}}

\def\nn{\nonumber}

\def\si{\sigma}

\def\de{\delta}

\def\La{\Lambda}
\def\eps{\epsilon}

\def\rd{{\rm d}}

\begin{document}

\vspace{1cm}

\centerline{\bf\Large Horizons and the cosmological constant}

\vspace{5mm}
\centerline{\bf Krzysztof A. Meissner}

\vspace{5mm}
\begin{center}
{\it Institute of Theoretical Physics, Dept. of Physics,\\
University of Warsaw, Ho\.za 69, 00-681 Warsaw, Poland}
\end{center}

\begin{abstract}

\footnotesize{\noindent A new solution of the Einstein equations
for the point mass immersed in the de Sitter Universe is presented.
The properties of the metric are very different from both the
Schwarzschild black hole and the de Sitter Universe: it is
everywhere smooth, light can propagate outward through the
horizon, there is an antitrapped surface enclosing the point mass
and there is necessarily an initial singularity. The solution for
any positive cosmological constant is qualitatively different from the Schwarzschild
solution and is not its continuous deformation.}
\end{abstract}

\vspace{5mm}\noindent{\bf 1. Introduction}

\vspace{2mm}
There is an extensive literature on the solutions of the Einstein
equations describing a point mass immersed in the expanding Universe
(see \cite{AK} and the recent review \cite{NG} and references
therein). The first exact solution for a point mass in the de Sitter
Universe was constructed as early as 1918 and is known as the
Kottler metric \cite{FK}:
\beq
\rd s^2= -\left(1-H^2 r^2 -\frac{2GM}{r}\right)\rd t^2 +
\frac{\rd r^2}{\displaystyle{1-H^2 r^2-\frac{2GM}{r}}}+r^2\rd\Omega^2
\label{Kott}
\eeq
where $H$ is related to the cosmological constant $\La$ by
$\La=3H^2/(8\pi G)$. For $H=0$ the metric is equal to the standard
Schwarzschild metric \cite{KS}
\beq
\rd s^2= -\left(1-\frac{2GM}{r}\right)\rd t^2 +
\frac{\rd r^2}{\displaystyle{1-\frac{2GM}{r}}}+r^2\rd\Omega^2
\eeq
However, for $M=0$ the metric requires a diffeomorphism to transform
it to the usual de Sitter metric \cite{dS}
\beq
\rd s^2= -\rd t^2 + e^{2Ht}\left(\rd r^2+r^2\rd\Omega^2\right)
\eeq
It is the purpose of this paper to present a new solution that in
the above mentioned limits goes over to the standard Schwarzschild
(at least outside of the black hole horizon) or de Sitter metrics.
We will show that in the limit of small cosmological constant the
metric indeed goes to the Schwarzschild metric but only outside the
Schwarzschild horizon -- inside it is everywhere singular.

\vspace{5mm}
\noindent{\bf 2. The metric}

\vspace{2mm}
The metric that solves
\beq
R_{\mu\nu} -\frac12 g_{\mu\nu}R=-8\pi G\La g_{\mu\nu}
\eeq
with required properties reads
\beq
\rd s^2= -f(t,r)\rd t^2 +
\frac{e^{2Ht}\rd r^2}{f(t,r)}+e^{2Ht}r^2\rd\Omega^2
\label{soln}
\eeq
where $\La=3H^2/(8\pi G)$,
\beq
f(t,r)=h(t,r)+\sqrt{h(t,r)^2+H^2 r^2 e^{2Ht}}
\eeq
and
\beq
h(t,r)=\frac12\left(1-H^2r^2e^{2Ht}-\frac{r_S}{r e^{Ht}}\right)
\eeq
where $r_S=2GM$. We see that if $H\ne 0$ (however small) $f(t,r)$
never vanishes. It is important to note that in the limit $H\to 0$
the solution tends to the Schwarzschild metric only outside of the
horizon and is undefined inside. The solution is connected to
(\ref{Kott}) by a singular diffeomorphism transformation so they
cover different patches of spacetime.

For some purposes it may be useful to write the metric in
coordinates $(t,\rho)=(t,r e^{Ht})$ (the metric is then independent
of time):
\beq
\rd s^2= -2h(0,\rho)\rd t^2 -\frac{2H\rho\rd t\rd\rho}{f(0,\rho)}+
\frac{\rd \rho^2}{f(0,\rho)}+\rho^2\rd\Omega^2
\eeq
with the inverse metric in these coordinates
\beq
g^{tt}=-\frac{1}{f(0,\rho)},\
\ g^{t\rho}=-\frac{H\rho}{f(0,\rho)},\
\ g^{\rho\rho}=2 h(0,\rho)
\eeq
This form shows that there are 4 Killing vectors (1 connected with
time translations and 3 with rotations) as in the Schwarzschild
case. In general there are two radii $\rho_1$, $\rho_2$ solving the
equation
\beq
h(0,\rho_i)=0
\eeq
and defining two null surfaces. The second null vector is in both
cases directed outside the surfaces therefore the inner horizon is of 
different type than the usual black hole horizon.

\vspace{5mm}
\noindent{\bf 3. Singularities}

\vspace{2mm}
We calculate the Riemann tensor for the metric (\ref{soln})
\bea
R^{01}{}_{01}\!&=&\! R^{23}{}_{23}=H^2+\frac{r_S}{r^3e^{3Ht}}\nn\\
R^{02}{}_{02}\!&=&\! R^{03}{}_{03}=R^{12}{}_{12}=R^{13}{}_{13}=
H^2-\frac{r_S}{2r^3e^{3Ht}}
\eea
with all other  components (except with permuted indices) vanishing.
Therefore the Ricci tensor and the curvature scalar read
\beq
R^\mu{}_\nu=3H^2\de^\mu_\nu,\ \ \ R=12H^2
\eeq
The only singularity of the scalars constructed from the Riemann
tensor like $R^{\mu\nu}{}_{\rho\si}R^{\rho\si}{}_{\mu\nu}$ occurs
for $r\to 0$ (as is the case also for the Schwarzschild solution) or
for $t\to -\infty$.

We now analyze the metric (\ref{soln}). For fixed $t$
\beq
f(t,r)\ \stackrel{r\to\infty}{\longrightarrow}\ 1,\ \ \ f(t,r)\
\stackrel{r\to 0}{\longrightarrow}\
\frac{H^2 r^3 e^{3Ht}}{r_S-r e^{Ht}}
\label{rlim}
\eeq
while for fixed $r$
\beq
f(t,r)\ \stackrel{t\to\infty}{\longrightarrow}\ 1,\ \ \ f(t,r)\
\stackrel{t\to -\infty}{\longrightarrow}\
\frac{H^2 r^3 e^{3Ht}}{r_S-r e^{Ht}}
\label{tlim}
\eeq
Therefore the physical radial distance to $r\to\infty$ is infinite
as well as the physical time interval to $t\to \infty$. The physical
radial distance to the origin $r=0$ is also infinite (although the
area is $4\pi r^2 e^{2Ht}$ i.e finite and small)
\beq
\int_\eps \frac{e^{Ht}\rd r}{\sqrt{f(t,r)}}
\sim \frac12 \sqrt{\frac{r_S}{H^2 \eps e^{Ht}}}
\eeq
so the space is geodesically complete. However the physical time
interval from $t\to -\infty$ to some $T$ (within the applicability
of (\ref{tlim}))
\beq
\int_{-\infty}^T \rd t\ \sqrt{f(t,r)}
\sim \frac23 \sqrt{\frac{r^3 e^{3HT}}{r_S}}
\eeq
is finite. Therefore in the presence of a point mass and the
cosmological constant there must be an initial singularity in
contradistinction to the usual de Sitter Universe.

The topology of the spacetime described by (\ref{soln}) is
$\mathbb{R}^2\times S^2$.

\vspace{5mm}
\noindent{\bf 4. Propagation of light}

\vspace{2mm}
Since the $g_{\mu\nu}$ components are nowhere vanishing and nowhere
singular the coordinates seem to cover the whole spacetime. To check
what happens at the inner horizon (corresponding to $r=2GM$
in the Schwarzschild case) we have to find the behaviour of light
both inside and outside. To do it we solve the equation for the
propagation of light i.e. stemming from $\rd s^2=0$ and directed
radially outwards i.e. satisfying
\beq
e^{Ht}\frac{\rd r(t)}{\rd t}=f(t,r(t))
\eeq
To simplify the discussion we assume in what follows that
\beq
r_S H\ll 1
\eeq
what is extremely well satisfied for any object in the Universe.
Then in these coordinates the horizon is given by
\beq
\frac{r_S}{re^{Ht}}\approx 1
\label{wbh}
\eeq
We have to distinguish two cases -- the initial position at $t=0$
$r_0<r_S$ or $r_0>r_S$.

In the latter case ($r_0>r_S$) the solution reads
\beq
r(t)=r_0+\frac{1}{H}\left(1-e^{-Ht}\right) +r_S\ln\frac{r_0 H}{1+r_0
H-e^{-Ht}}+O(r_S^2 H)
\label{traj}
\eeq
so that after infinite time the comoving coordinate reaches the de
Sitter horizon
\beq
r(\infty)\approx r_0+r_S\ln(r_0 H)+\frac{1}{H}
\eeq

A different situation arises if $r_0<r_S$. Then $f(t,r)$ is positive
but extremely small so that $r(t)$ increases very slowly until it
reaches the horizon and time grows to the value
corresponding to (\ref{wbh}) i.e.
\beq
T_0=\frac{1}{H}\ln\left(\frac{r_S}{r_0}\right)
\eeq
Then the light gets outside and the trajectory is given by
(\ref{traj}) but with $t\to t+T_0$. Therefore the total comoving
distance is approximately given by
\beq
r(\infty)\approx \frac{r_0}{r_S H}
\eeq
i.e. less than the outside de Sitter horizon by a factor
$\frac{r_0}{r_S}$.

It is important to check the dependence on time of the area enclosed
by the outgoing or incoming light
\beq
\frac{\rd A}{\rd t}=8\pi r e^{2Ht}
\left(Hr\pm f(t,r) e^{-Ht}\right)
\eeq
where $A=4\pi r^2 e^{2Ht}$. Since $Hr>f(t,r) e^{-Ht}$ for
sufficiently small $r$ it turns out that there is an antitrapped
surface enclosing some region around $r=0$. This is consistent with
the fact that antitrapped surfaces must have a singularity in the
past as we have seen is the case for the metric (\ref{soln}).

\vspace{5mm}\noindent{\bf 4. Conclusions}

\vspace{2mm}

In the paper we have shown that the solution of the Einstein
equations (\ref{soln}) for a point mass immersed in the universe
with the positive cosmological constant has very special properties:
the metric is everywhere smooth, light can propagate outward through
the horizon, there is an antitrapped surface enclosing the
point mass and there is necessarily an initial singularity. Although
with extremely small value of $H$ such an object for all practical
purposes looks like a usual black hole the conceptual difference
resulting from the fact that there is no horizon for the outward
propagation of light can be far-reaching -- first, one should
rethink a notion of a black hole entropy as proportional to the area
of the horizon and second, there seems to be no information loss
even classically since the communication of the inside with the
outside is extremely weak but nonvanishing. It is also interesting
to note that in the presence of such objects there is necessarily an
initial singularity in distinction to the pure de Sitter universe
and there is no continuous deformation connecting $\La>0$ solution
described in this paper and the Schwarzschild metric.

\vspace{3mm}\noindent{\bf Acknowledgements}
The author is grateful to D. Christodoulou, J. Lewandowski, H.
Nicolai and A. Trautman for discussions and to Max Planck Institute
in Potsdam for hospitality. The work was partially supported by the
EU grant MRTN-CT-2006-035863 and the Polish grant N202 081 32/1844.


\begin{thebibliography}{Ref}

\bibitem{AK} A. Krasi{$\acute{\rm n}$}ski,
{\it Inhomogeneous Cosmological Models},
Cambridge Monographs on Mathematical Physics 1998 (Cambridge
University Press, Cambridge).

\bibitem{NG}
M. Carrera, D. Giulini, {\it On the influence of global cosmological
expansion on the dynamics and kinematics of local systems}, {\tt
arXiv:0810.2712 [gr-qc]}

\bibitem{FK} F. Kottler, Ann. Phys. (Leipzig) 56 (1918) 401.

\bibitem{KS} K. Schwarzschild, {\it {\"U}ber das Gravitationsfeld
eines Massenpunktes nach der Einsteinschen Theorie},
Sitzungsberichte der K{\"o}niglich-Preussischen Akademie der
Wissenschaften, Berlin (1916) pp 189.

\bibitem{dS} W. de Sitter,  {\it On the relativity of inertia: Remarks
concerning Einstein's latest hypothesis}, Proc. Kon. Ned. Acad. Wet.
19: (1917) 1217-1225; {\it On the curvature of space}, ibid. 20:
(1917) 229-243.
\end{thebibliography}
\end{document}